\documentclass[amsmath,amssymb,prl]{revtex4}
\usepackage{graphicx}

\newcommand{\beq}{\begin{equation}}
\newcommand{\eeq}{\end{equation}}
\newcommand{\beqa}{\begin{eqnarray}}
\newcommand{\eeqa}{\end{eqnarray}}
\newcommand{\non}{\nonumber}

\newcommand{\En}{ E_i^{(N)}({\bf m}) }

\newcommand{\mv}{{\bf m}}

\newcommand{\hv}{{\bf h}}
\newcommand{\fv}{{\bf f}}

\newcommand{\Nun}{d{\cal N}^{(N)}}
\newcommand{\Nunn}{d{\cal N}^{(N+1)}}

\begin{document}

\title{Cavity Method for Supersymmetry Breaking Spin Glasses}

\author{Andrea Cavagna$^{\dagger}$, Irene Giardina$^{\dagger}$ and 
Giorgio Parisi$^{\dagger \ddagger}$}

\affiliation{$^{\dagger}$ Dipartimento di Fisica, Universit\`a di Roma ``La Sapienza''  and
Center for Statistical Mechanics and Complexity, INFM Roma 1 
P.le Aldo Moro 2, 00185 Roma,  Italy
\\
$^{\ddagger}$ Istituto Nazionale di Fisica Nucleare, Sezione Roma 1}

\date{July 16, 2004}

\begin{abstract}
The spontaneous supersymmetry-breaking that takes place in certain spin-glass models signals a
particular fragility in the structure of metastable states of such systems.  This fragility is due
to the presence of at least one marginal mode in the Hessian of the free energy, that makes the
states highly susceptible under external perturbations.  The cavity method is a technique that
recursively describes the property of a system with $N+1$ spins in terms of those of a system with
$N$ spins.  To do so, the cavity method assumes a certain degree of stability when adding a new spin
to the system,  i.e. it assumes that for a generic choice of the parameters there is an one-to-one
correspondence between the metastable states of the system with $N$ spins and the metastable states of
the system with $N+1$ spins.  In systems where the supersymmetry is broken such a correspondence does
not exist, and an alternative formulation of the cavity
method must be devised.  We introduce a generalized cavity approach that takes care of this problem
and we apply it to the computation of the probability distribution of the local magnetizations in
the Sherrington-Kirkpatrick model.  Our findings agree with the correct supersymmetry-breaking
result.
\end{abstract}

\maketitle

In physics we often meet two quite different situations. 
On the one hand, we tend to assume (or hope) that small perturbations have small effects. 
This is at the basis of all attempts to use any sort of perturbation theory.
On the other hand, we know that many interesting phenomena take place  
when such an assumption is in fact violated. At
criticality, whenever the susceptibility of a system is infinite, its response is anomalous. In this case
small perturbations may indeed have  big effects. This state of affairs is often accompanied by the 
presence of zero, or marginal, modes among the second derivatives of the action, may these be the masses, 
or the inverse susceptibilities. When this happens we must be extremely careful in applying 
perturbation theory. 

A particularly interesting realization of this scenario occurs when criticality is induced by
spontaneous symmetry breaking.  In that case, symmetry breaking, marginality, and break-down of
standard perturbative techniques, give rise to some interesting phenomena.  In this work we
discuss how the spontaneous breaking of a supersymmetry is connected to the presence of marginal
modes and anomalous response in certain spin-glass models.  The main focus of our study is the
structure of metastable states in such models, and in particular the extreme fragility of this
structure under small perturbations.  A significant consequence of this feature is the break-down of
the standard cavity method \cite{cavity,mpv}, whose basic assumption is that by adding one extra
degree of freedom to a large systems, its physical properties (and in particular some properties of
the structure of states) do not change dramatically.  This assumption is no longer valid when the
supersymmetry is broken.  Our main result is to provide a generalization of the cavity method that
works also in supersymmetry-breaking systems.

Metastable states in mean-field spin-glasses can be identified with the local minima of 
a mean-field free energy $F$ (also known as the TAP free energy \cite{tap}), that is a function of 
the local magnetizations $m_i$ of the system.  Generally speaking, in order to compute 
the number of local minima of a function $F(\mv)$ , 
one can introduce an effective action that is invariant under a generalized form of the 
Becchi-Rouet-Stora-Tyutin (BRST) supersymmetry \cite{brs,sourlas,juanpe,brst1,leuzziprb}. 
The Ward identities generated by this symmetry have a rather clear physical meaning. 
The most relevant one reads  \cite{leuzziprb},
\beq
\left.\frac{d\langle  m_i \rangle}{dh_j}\right|_{h=0} =  \langle A_{ij}(\mv) \rangle \ ,
\label{ward}
\eeq where $A_{ij}(\mv) = [\partial_i\partial_j F(\mv)]^{-1}$ is the inverse of the second
derivative, i.e. of the Hessian.  The brackets $\langle\cdot\rangle$ indicate a sum over all metastable
states of the system.  The meaning of equation (\ref{ward}) is straightforward: it expresses the
natural relationship between the susceptibility and the curvature of the minima of the free energy.
Therefore, the supersymmetry seems to encode a very robust physical feature of metastable states,
namely the static fluctuation-dissipation theorem.  This notwithstanding, it is now believed that in
certain spin-glass models the supersymmetry is in fact spontaneously broken in the low-temperature
phase, and that relation (\ref{ward}) is thus violated \cite{abm,brstprl,gold}.  Such systems
include the Sherrington-Kirkpatrick (SK) model \cite{sk}, that we discuss here.

The reason why the static fluctuation-dissipation theorem is violated, and thus the supersymmetry is
broken, lies in the peculiar structure of metastable states of such systems.  Recent studies show
\cite{abm,brstprl,gold} that at low temperatures {\it all} stationary points of the mean-field free
energy are organized into minimum-saddle pairs.  The minimum and the saddle are connected along a
mode that is softer the larger the system size $N$.  Moreover, the free energy difference of the
paired stationary points decreases with increasing $N$.  In other words, supersymmetry-breaking
metastable states are marginal in the thermodynamic limit, having at least one flat zero mode.  In
this situation it is clear that even an infinitesimal external field may destabilize some states,
making them disappear.  On the other hand, virtual states, i.e. inflection points of the free energy
with a very small second derivative, may be stabilized by the field, giving rise to pairs of new
states.  In such a situation we must reconsider the validity of equation (\ref{ward}).  At the
l.h.s. we differentiate with respect to an external field a sum over all metastable states.  The
problem is that, due to marginality, {\it some elements in this sum may disappear or appear as the
field goes to zero}.  Therefore, even though the static fluctuation-dissipation relation holds for
each individual state, when we sum over all states, an anomalous contribution arises due to the
instability of the whole structure with respect to the field.  Relation (\ref{ward}) is thus
violated.

Supersymmetry breaking is thus the mathematical expression of a great instability in the structure
of metastable states.  Adding a new spin to the system, the basic building block of the cavity
method, will therefore not be harmless.  Our aim is to find a more general formulation of the
method, valid also in supersymmetry-breaking systems.  Note that a similar task has been independently
pursued in \cite{rizzo}, although introducing an explicit violation of the supersymmetry.  Our starting
point is the set of equations satisfied by the local magnetizations in a system with $N$ degrees of
freedom.  These are just the equations of stationarity of the free energy, 
\beq \frac{\partial
F}{\partial m_i} \equiv E_i^{(N)}(\, \mv)= 0 \hskip 1 truecm i=1,\dots,N \ ,
\label{eqN}
\eeq
where $\mv= m_1,\dots,m_N$. For the SK model we have \cite{tap},
\beq
E_i^{(N)} \ \equiv\  \tanh^{-1} m_i +\beta^2(1-q)\, m_i-\beta\sum_{j=1}^N J_{ij} m_j  \ .
\label{tap} 
\eeq
The self-overlap is $q=(1/N)\sum_i m_i^2$, and the random couplings $J_{ij}$ are drawn from a Gaussian 
distribution of mean zero and variance $1/N$.
Let us introduce the number $\Nun(\mv)$ of solutions of eqs.(\ref{eqN}) with magnetizations between $\mv$ and 
$\mv +d\mv$. 
The density of solutions $\rho^{(N)}(\mv)$ is then defined as,
\beq
\Nun(\mv) =    \rho^{(N)}(\mv) \ d \mv \  .
\eeq
The basic point of the cavity method \cite{cavity,mpv} is to write a recursive relation expressing
the density $\rho^{(N+1)}$ of the system with $N+1$ spin, as a function of the density $\rho^{(N)}$
of the system with $N$ spins.  The fundamental idea behind this procedure is that adding one single
spin to a system with a large number of degrees of freedom, is a small perturbation and consequently
the properties of the new $N+1$ system will not be much different from the original $N$ system.
More specifically, a natural assumption is that the structure of metastable states of the system
does not change when we add the new spin: the solutions of (\ref{eqN}) are slightly modified,
but there is still a one-to-one correspondence between solutions of the $N$ and $N+1$ systems.
This hypothesis seems reasonable, and it has been at the basis of all standard cavity investigations.  
However, we will show that it is in fact equivalent to assume that the supersymmetry
is unbroken.

Let us add a new spin at site $0$, and call $(\mv,m_0)$ the global magnetization vector of the $N+1$
system.  The new set of $N+1$ equations can be split as,
\beqa
E_i^{(N+1)}(\mv,m_0) &=& 0 \hskip 1 truecm i=1,\dots,N \ ,
\label{eqNN}\\
E_0(\mv,m_0)&=& 0 \ .
\label{eqzz}
\eeqa
We can examine this  set of equations by first solving the first $N$ equations 
(\ref{eqNN}) at fixed $m_0$, and then plugging $\mv(m_0)$ into (\ref{eqzz}). 
The assumption of stability of states can be expressed by saying that there is a one-to-one mapping  
between solutions $\mv^{(N+1)}$ of eqs.(\ref{eqNN}) and solutions $\mv^{(N)}$ of eqs.(\ref{eqN}):
if $\mv^{(N)}$ is a solution  of the $N$ system, the effect of adding the spin is just 
to slightly modify the old magnetization to a new value  
$\mv^{(N+1)}={\bf f}_{m_0}(\mv^{(N)})$. 
From this key hypothesis, it follows that for a given $m_0$ the number of solutions with magnetization 
$\mv^{(N+1)}$ of the first $N$ equations of the 
$N+1$ system, is simply equal to the number of solutions of the original $N$ system that have 
magnetization $\mv^{(N)}$,  
\beq
\Nunn(\mv^{(N+1)})=\Nun(\mv^{(N)}) \ ,
\label{basic}
\eeq
with $\mv^{(N+1)}={\bf f}_{m_0}(\mv^{(N)})$.
In order to obtain the density of solutions of the new $N+1$ system we must also impose the 
extra equation (\ref{eqzz}) for $m_0$. This relation can be cast in a simple physical form, namely
$m_0 = \tanh\,(\beta \sum_j J_{0j} m_j^{(N)})$:
the spin at zero aligns to the local field produced on site $0$ by the original $N$ system. 
The density of solutions for the new system therefore becomes,
(sums over repeated indices are understood),
\beq
   \rho^{(N+1)}\left(\mv^{(N+1)},m_0\right) \, d\mv^{(N+1)} \, dm_0
=  \rho^{(N)}\left(\mv^{(N)}\right) \, d\mv^{(N)} \ \ \delta \left( \tanh^{-1} m_0 - \beta  J_{0j} m_j^{(N)} \right) 
\ \frac{1}{1-m_0^2} \ dm_0  \ ,
\label{ninetta}
\eeq
The first part of this equation is equivalent to  relation (\ref{basic}), while the second part makes sure that the
equation for $m_0$ is satisfied (the last factor is the Jacobian of the $\delta$-function). The meaning of this equation
is clear: the density of solutions remains the same, provided that we satisfy the equation for the new site $0$, and that
we change a little the other $N$ magnetizations. The map between the old and the new magnetizations 
can be found by noting that the new magnetization $m_0$ exerts a field $J_{i0} m_0$ on the old sites, and thus
$m_i^{(N+1)}=m_i^{(N)}+\chi \,J_{i0}\, m_0$, where $\chi=\beta(1-q)$ is the susceptibility.

Equation $(\ref{ninetta})$ is a recursive relation for the number of states. In order to get a self-consistency equation, 
we have to work with the probability, i.e. the density divided by the total number of states, $\rho(\mv)/{\cal N}$, that 
has a well defined limit when $N\to\infty$. In the annealed approximation we assume that the probability is self-averaging
and that it can therefore be averaged over the disorder \cite{nota1}. 
The averaged probability density  factorizes, i.e.   $\overline{\rho(\mv)/{\cal N}}=  \prod_i\, p(m_i)$, and thus
Eq. (\ref{ninetta}) becomes a self-consistency equation 
for the single site probability  $p(m_i)$: we divide both sides by the global number
of states ${\cal N}^{(N+1)}$, average   over the disorder, and integrate over the first $N$ magnetizations. 
In the thermodynamic limit $p^{(N+1)}=p^{(N)}$, and we  finally get,
\beq
p(m_0) = \int \prod_i dP(J_{i0})\ dm_i \ p(m_i) \ {\cal K}_{\rm ss}(\mv,m_0) \ ,
\label{susy}
\eeq
where,
\beq
{\cal K}_{\rm ss}(\mv,m_0) =
\delta\left(\tanh^{-1}m_0 - \beta J_{0j} m_j\right)/(1-m_0^2) \ .
\eeq
The factor ${\cal N}^{(N)}/{\cal N}^{(N+1)}$ on the r.h.s. has been reabsorbed into the normalization constant
of the probability $p$. This equation is not difficult to solve, and we find,
\beq
p(m_0) = \frac{1}{\sqrt{2\pi\beta^2 q}} \exp\left[-\frac{ (\tanh^{-1} m_0)^2}{2\beta^2 q}\right] \ \frac{1}{1-m_0^2} \ ,
\label{psusy}
\eeq
where $q=\langle m^2\rangle$, and where  $\langle X \rangle = \int dm\, p(m)\, X(m)$.
Equation (\ref{psusy}) coincides with the probability 
distribution found in \cite{brst1}, that is the {\it supersymmetric} distribution.
The assumption of stability of metastable states is thus equivalent 
to assuming unbroken supersymmetry. However,  the BRST supersymmetry is in fact broken 
in the SK model \cite{abm,brstprl},
and thus the correct distribution is not given by (\ref{psusy}). To find the  supersymmetry-breaking distribution
within the cavity approach,  we must give up the assumption of stability of states, and find a more general formulation.
The first step is to recognize that eqs.(\ref{eqNN}) for the $N+1$ system 
can be formally rewritten as,
\beq
E^{(N+1)}_i(\mv,m_0 | \beta) = E^{(N)}_i(\mv |\beta') - k_i(\mv,m_0) =0
\label{link}
\eeq
with,
\beq
k_i(\mv,m_0)=\beta J_{i0} m_0 - \frac{\beta^2}{N}(1-m_0^2)\, m_i \ .
\eeq
The variance of  $J_{ij}$ in the $N+1$ system   is smaller than in the $N$ system. 
We have then rescaled the inverse temperature from $\beta$ to  $\beta'=\beta [1 - 1/(2N)]$ in such a way that
the disorder appearing in $E_i^{(N)}$ has the correct scaling for a system of size $N$.
Equation (\ref{link}) suggests the path we have to follow: the
function $k_i$ may be seen as a local field acting on site $i$ of the $N$ system, so that, in a way, finding
solutions of the $N+1$ system is like finding solutions of the $N$ system with a field. If this can be done,
we may hope to write a self-consistency  equation not simply for the probability density of solutions with given 
magnetization, but for a more complicated object, i.e. the probability density of solutions with given magnetization 
{\it and} field. The equations for the metastable states of an $N$ system with a field are,
\beq
\En=h_i \ .
\label{nini}
\eeq
We define the new density of solutions  $\rho^{(N)}(\mv|\hv)$  as,
\beq
d{\cal N}^{(N)}(\mv|\hv) = \rho^{(N)}(\mv|\hv)\ d\mv \ ,
\eeq
where $d{\cal N}^{(N)}(\mv|\hv)$ is the number of solutions of (\ref{nini}), with given magnetization $\mv$ and 
{\it external} field ${\bf h}$. The idea is to write a recursive relation for $\rho^{(N)}(\mv|\hv)$ rather than 
for $\rho^{(N)}(\mv)$. From (\ref{link}) we have that the equations for the $N+1$ system in
an external field $(\hv,h_0)$ are,
\beqa
E^{(N)}_i(\mv) &=& k_i(\mv,m_0) + h_i 
\label{botto}
\\
E_0(\mv,m_0) &=& h_0 \ .
\label{cotto}
\eeqa
Let us  focus now on (\ref{botto}), at fixed value of $m_0$. The number of solutions 
of this set of equations seems formally equal to the number of solutions of a system with external field 
${\bf k} + \hv$. The only problem is that ${\bf k}$ is {\it not} an external field, since it depends on 
$\mv$, and thus $d{\cal N}(\mv|\hv+{\bf k}(\mv)) \neq  \rho(\mv|\hv+{\bf k}(\mv))\, d\mv$. 
Fortunately,  it can be simply proved a similar relation, that reads,
\beq 
d{\cal N}^{(N)}(\mv|\hv+{\bf k}(\mv))  =\rho^{(N)}(\mv|\hv+{\bf k}(\mv)) \ \omega(\mv)\ d\mv \ ,
\label{nuova}
\eeq
where the re-weighting factor is given by,
\beq
\omega(\mv)=  \ \frac{\partial_{\mv}[E^{(N)}_i(\mv) - k_i(\mv)]}{\partial_{\mv}[E^{(N)}_i(\mv)]} \ .
\eeq
We can  apply this general formula to equation (\ref{botto}), and remember that (as in the previous case) to
obtain the full solutions density for the $N+1$ system we must also satisfy the extra equation for $m_0$. This time, 
however, there is no mapping between solutions of the $N$ and $N+1$ systems, and thus the field experienced by spin $0$
must be expressed as a function of the new magnetizations \cite{mpv},
\beq
E_0= \tanh^{-1}m_0 - \beta\sum_j J_{0j} m_j + \beta^2(1-q)\, m_0 \ ,
\eeq
and we have to remember that $\mv$ depends on $m_0$. In conclusion, we have,
\beqa
\rho^{(N+1)}(\mv,m_0|\hv,h_0) &=& 
\rho^{(N)}(\mv|\hv+{\bf k}(\mv,m_0)) \  \omega(\mv,m_0) \times
\non \\
&&\phantom{swissphantomimegiga}
\times \ \delta\left(E_0(\mv,m_0) - h_0 \right) \ \frac{dE_0}{dm_0}  \ .
\label{rino}
\eeqa
Equation (\ref{rino}) is a general recursive relation for the density of states, that is {\it always} valid,
whether the structure of states is stable or not. The supersymmetric result can be recovered if we assume that 
there is a mapping between solutions $\mv$ of the $N+1$ system and solutions $ \fv^{-1}_{m_0}(\mv)$ of the
system in absence of the $0$ site. 
In this case we have (for any fixed $N$),
\beq
\rho(\mv|\hv+{\bf k}(\mv,m_0)) \ \omega(\mv,m_0) \ d\mv  
= \rho(\fv^{-1}_{m_0}(\mv)|\hv) \ d\fv^{-1}_{m_0} .
\eeq 
In other words, in the supersymmetric case the effective field ${\bf k}$ can be reabsorbed via a change of variables, 
giving back the supersymmetric expression (\ref{ninetta}).

When we pass from single instances to probabilities, we can average equation (\ref{rino}) over the disorder and once again 
write self-consistency  equation for the single-site probability $p(m_i|h_i)$. After some algebra we get,
\beq
p_\beta(m_0|h_0) = 
\int \prod_i dP(J_{i0}) \ dm_i \ p_{\beta'}(m_i|h_i+k_i(m_i,m_0)) \ {\cal K}(\mv,m_0)
\label{p}
\eeq
with,
\beq
{\cal K}(\mv,m_0)=
 \delta\left[\tanh^{-1}m_0 - \beta J_{0j}m_j  + \beta^2(1-q)m_0 -h_0\right] 
\ \exp[\beta^2(1-m_0^2)(1-q)]\ \frac{1}{1-m_0^2} \ ,
\label{rutto}
\eeq
and where we recall that, $\beta'=\beta [1 - 1/(2N)]$. The exponential in (\ref{rutto}) is the factor $\omega$, while
the last term is the Jacobian of the delta.
Let us compare equation (\ref{susy}) to (\ref{p}):
the key point  is that in the general case there is no closed equation for 
$p(m_i)$, however complicated. On the other hand,  if we start from a system with external magnetic field, the effect of the new spin can be counterbalanced by tuning the local fields in the new system, in such a way to leave stable the structure of states. For this reason it is  possible  to write a self-consistent equation only for the probability $p(m_i|h_i)$, but {\it not} for $p(m_i)$. This is the core of  our approach.

In order to solve equation (\ref{p}) we adopt an ansatz for the asymptotic form of $p(m|h)$ (we drop now the 
subscript $0$). For small external fields, we expect,
\beq
p(m|h) =  p(m)\  \exp\left[A(m) h^2 +B(m) h\right] \ ,
\label{ansatz}
\eeq
where terms of higher order in $h$ have been discarded, and $A(m)$ and $B(m)$ are functions to be determined 
self-consistently.  By setting $h_i=0$ and $h_0\sim k_i \sim 1/\sqrt{N}$ in equation (\ref{p}), and 
by using an integral representation for the delta function, we get to leading order in $N$:
\beqa
p(m|h)
&=& 
C 
\exp\left\{  -\frac{1}{2\beta^2 q}[\tanh^{-1}m-\Delta m]^2 +\lambda m^2\right\}\, \frac{1}{(1-m)^2}
\times \non\\
&&\phantom{titolivionazi}\times 
\exp\left\{ -\frac{1}{2\beta^2 q}\,h^2 +\frac{1}{\beta^2 q}[\tanh^{-1}m-\Delta m]\,h\right\} \ .
\label{mendieta}
\eeqa
with  $q=\langle m^2\rangle$, $A=\langle A(m)\rangle$, 
$\Delta =\beta^2 \langle mB(m)\rangle - \beta^2 (1-q)$ 
and $\lambda = (\beta^2/2) [ \langle B(m)^2\rangle + A]+ \Delta$.
We note that the shift in temperature only gives  a constant term at the leading order: 
together with all other constant contributions it has been reabsorbed in the normalization constant $C$. 
By comparing (\ref{ansatz}) with (\ref{mendieta}) we immediately have,
$ A(m) = -1/(2 \beta^2 q)$,  $B(m) = [\tanh^{-1}m-\Delta m]/(\beta^2 q)$.
The parameters $\Delta$ and $\lambda$ therefore satisfy the equations:
\beqa
\Delta &=& \frac{1}{2q}\langle m\,\tanh^{-1}m \rangle -\frac{\beta^2}{2}(1-q)   
\non \\
\lambda &=&\Delta -\frac{1}{2q} +\frac{1}{2\beta^2 q^2} \langle[\tanh^{-1} m-\Delta m]^2\rangle \ .
\eeqa
The expression of $p(m)$ we have found (first line in equation (\ref{mendieta})), 
and the equations for the parameters $\Delta$ and $\lambda$ 
are precisely those of the supersymmetry-breaking solution  found long ago in \cite{bm1}. 
The supersymmetric expression is recovered for $\Delta=\lambda=0$ \cite{brst1}.

In this work we have shown how the cavity method can be generalized to situations where the structure
of states is unstable under external perturbations, i.e. where the supersymmetry is broken. 
In this case, the simple distribution $p(m)$ is not stable when a
new spin is added to the system. However, the perturbation caused
by the new spin is analogous to an external field. Therefore, if we consider the system in presence
of a field, we can balance the effect of the new spin by tuning the field.
This means that the distribution  $p(m|h)$ is stable when the new spin is added, and a self-consistent relation 
can be written for it. It is important to stress that this equation  can be written for any system, once the
form of the effective field $k_i$ is known. In particular, applications of this method 
to diluted systems are under study. 
We thank F. Ricci Tersenghi for many interesting discussions.


\begin{thebibliography}{22}

\bibitem{cavity} M.Mezard, G.Parisi and M.A. Virasoro, Europhys. Lett. {\bf 1}, 77 (1986).

\bibitem{mpv} M.Mezard, G.Parisi and M.A. Virasoro, {\it Spin Glass Theory and beyond}, World Scientific, 
Singapore (1987).

\bibitem{tap} D.J. Thouless, P.W. Anderson and R.G. Palmer, Phil. Mag., {\bf 35}, 593 (1977).

\bibitem{brs} C. Becchi, R. Rouet and A. Stora, Comm. Math. Phys. {\bf 42}, 127 (1975);
I.V. Tyutin {\it Lebedev preprint} FIAN 39 (1975). 

\bibitem{sourlas} G. Parisi and N. Sourlas, Phys. Rev. Lett. {\bf 43}, 744 (1979).

\bibitem{juanpe} A. Cavagna, J.P. Garrahan and I. Giardina, J. Phys. A {\bf 32}, 711 (1998).

\bibitem{brst1} A. Cavagna, I. Giardina, G. Parisi  and M. Mezard, J. Phys. A {\bf 36}, 1175 (2003).

\bibitem{leuzziprb} A. Crisanti, L. Leuzzi, G. Parisi, and T. Rizzo, Phys. Rev. B {\bf 68}, 174401 (2003).

\bibitem{abm}  T. Aspelmeier,  A. J. Bray,  M. A. Moore,  Phys. Rev. Lett. {\bf 92}, 087203 (2004)

\bibitem{brstprl} A. Cavagna, I. Giardina and  G. Parisi,  Phys. Rev. Lett. {\bf 92}, 120603 (2004)

\bibitem{gold} G. Parisi and T. Rizzo, Preprint cond-mat/0401509.

\bibitem{sk} D. Sherrington  and S. Kirkpatrick, Phys. Rev. Lett. {\bf 32}, 1792 (1975).

\bibitem{rizzo} T. Rizzo, Preprint cond-mat/0403261; Preprint cond-mat/0404729.

\bibitem{nota1} This approximation is in fact exact in the present case, since we do not impose any constraint 
on the free energy. 

\bibitem{bm1} A.J. Bray and M.A. Moore, J. Phys. C {\bf 13}, L469 (1980).


\end{thebibliography}
\end{document}